\documentclass{article}
\usepackage{graphicx}

\begin{document}

\thispagestyle{empty}                                                          
                                                                               
\begin{center}                                                                 
\begin{tabular}{p{130mm}}                                                      
                                                                               
\begin{center}                                                                 
{\bf\Large                                                                     
LOCALIZED STATES (``qubits'')} \\                                                            
\vspace{5mm}                                                                   
                                
{\bf\Large ENTANGLEMENT AND DECOHERENCE}\\
\vspace{5mm}

{\bf\Large FROM WIGNER ZOO}\\

\vspace{1cm}

{\bf\Large Antonina N. Fedorova, Michael G. Zeitlin}\\

\vspace{1cm}

{\bf\large\it
IPME RAS, St.~Petersburg,
V.O. Bolshoj pr., 61, 199178, Russia}\\
{\bf\large\it e-mail: zeitlin@math.ipme.ru}\\
{\bf\large\it e-mail: anton@math.ipme.ru}\\
{\bf\large\it http://www.ipme.ru/zeitlin.html}\\
{\bf\large\it http://www.ipme.nw.ru/zeitlin.html}

\end{center}

\vspace{1cm}

\begin{abstract}
We present the application of the variational-wavelet approach to                  
the construction and analysis of solutions of Wigner/von Neumann/Moyal and related equations
corresponding
to the nonlinear (polynomial) dynamical problems. (Naive) deformation                     
quantization, the multiresolution representation (exact multiscale decompostion) 
and the variational approach are the key points. We        
construct the solutions via the high-localized nonlinear eigenmodes in the base of the          
compactly supported wavelets and the wavelet packets. We demonstrate the appearance of (stable)
localized patterns (waveletons) and consider entanglement and decoherence as possible applications.
\end{abstract} 

\vspace{10mm}

\begin{center}
{\large Presented at the Fifth Workshop on Mysteries, Puzzles and Paradoxes in  
Quantum Mechanics, }\\
{\large Gargnano, Garda Lake, September 2003}

\vspace{10mm}

{\large J. of Modern Optics, Volume 51, Numbers 6-7, 1105, 2004} 
\end{center}
\end{tabular}
\end{center}
\newpage

{\bf\Large                                                                     
Localized states (``qubits''), entanglement 
and decoherence from Wigner zoo} \\                                                            

\vspace{0.5cm}

{\bf\Large Antonina N. Fedorova, Michael G. Zeitlin}\\

\vspace{0.5cm}

\noindent
{\bf IPME RAS, St.~Petersburg,
V.O. Bolshoj pr., 61, 199178, Russia,}
{\bf zeitlin@math.ipme.ru},
{\bf http://www.ipme.ru/zeitlin.html},\\
{\bf http://www.ipme.nw.ru/zeitlin.html}

\vspace{0.5cm}

We present the application of the variational-wavelet analysis to                  
the calculations and analysis of the solutions of Wigner/von Neumann/Moyal and related equations
corresponding
to the nonlinear (polynomial) dynamical problems [1], [2]. (Naive) deformation                     
quantization, the multiresolution 
representations and the variational approach are the key points. We        
construct the solutions via the multiscale expansions in the 
high-localized nonlinear eigenmodes in the base of the          
compactly supported wavelets and the wavelet packets. We demonstrate the appearance of (stable)
localized patterns (waveletons) and consider entanglement and decoherence as possible applications. 
Our goals are some attempt of classification and the explicit numerical-analytical constructions
of the existing zoo of possible/realizable quantum states.
There is a hope on the understanding of relation between the structure of initial Hamiltonians and
the possible types of quantum states and the qualitative type of their behaviour.
Inside the full spectrum  there are at least four possibilities which are the most 
important from our point of view for possible realization of quntum-like computations:
localized states, chaotic-like or/and entangled patterns, 
localized (stable) patterns (waveletons). 
We consider the calculations of the Wigner functions
$W(p,q,t)$ (WF) corresponding
to the classical polynomial Hamiltonians $H(p,q,t)$ as the solution
of the Wigner-von Neumann equation:
\begin{eqnarray}
i\hbar\frac{\partial}{\partial t}W = H * W - W * H
\end{eqnarray}
and related Wigner-like equations, e.g. 
Wigner transform of master equation (Lindblad-like) describing the decoherence 
\begin{eqnarray}
\dot{W}=\{H,W\}+
\sum_{n\geq 1}\frac{\hbar^{2n}(-1)^n}{2^{2n}(2n+1)!}
\partial^{2n+1}_x U(x)\partial_p^{2n+1}W+
2\gamma\partial_p pW+D\partial^2_pW
\end{eqnarray}
We have constructed the following quantum states: 
localized (as a model for qubit), entangled patterns,
localized (stable) patterns (e.g., 
modeling the decoherence as a result of interaction with environment). 
The obtained solutions of these equations have the 
following mul\-ti\-sca\-le or mul\-ti\-re\-so\-lu\-ti\-on decomposition via 
nonlinear high\--lo\-ca\-li\-zed eigenmodes 
\begin{eqnarray}
&&W(t,x_1,x_2,\dots)=
\sum_{(i,j)\in Z^2}a_{ij}U^i\otimes V^j(t,x_1,x_2,\dots),\nonumber\\
&&V^j(t)=
V_N^{j,slow}(t)+\sum_{l\geq N}V^j_l(\omega_lt), \quad \omega_l\sim 2^l, \\
&&U^i(x_s)=
U_M^{i,slow}(x_s)+\sum_{m\geq M}U^i_m(k^{s}_mx_s), \quad k^{s}_m\sim 2^m,
 \nonumber
\end{eqnarray}
which correspond to the full multiresolution expansion in all underlying phase space 
or space-time scales.
The formulae (3) give the expansion into a slow part
and fast oscillating parts for arbitrary $N, M$.  So, we may move
from the coarse scales of resolution to the 
finest ones for obtaining more detailed information about the dynamical process.
In this way one obtains contributions to the full solution
from each scale of resolution or each time/space scale or from each nonlinear eigenmode.
It should be noted that such representations 
give the best possible localization
properties in the corresponding (phase)space/time coordinates. 
Formulae (3) do not use perturbation
techniques or linearization procedures.
The modeling demonstrates the appearance of different (stable) pattern formation from
high-localized coherent structures or chaotic/entangled behaviour.
Our (nonlinear) eigenmodes (Fig. 1) are more realistic for the modeling of 
nonlinear classical/quantum dynamical process  than the corresponding linear gaussian-like
coherent states. Here we mention only the best convergence properties of the expansions 
based on wavelet packets, which  realize the minimal Shannon entropy property
and the exponential control of convergence of expansions like (3).
As an example, waveleton (Fig. 3) corresponds to result of superselection
(einselection) after decoherence process started from entaglement state demonstrated by 
Fig. 2 and constructed from 
localized ``qubit-like'' states (Fig. 1) [2].
It should be noted that
we can control the type of behaviour on the level of the reduced algebraic system [2].

\begin{figure}[ht]
\centering
\includegraphics*[width=65mm]{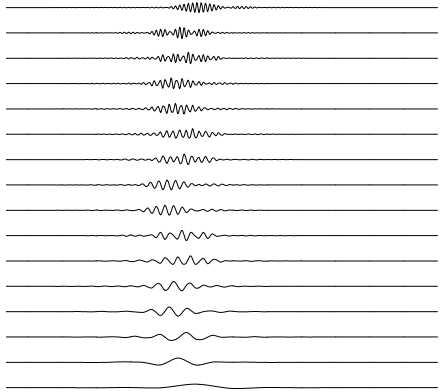}
\caption{\bf Localized modes.}
\end{figure}

\begin{figure}[ht]
\centering
\includegraphics*[width=65mm]{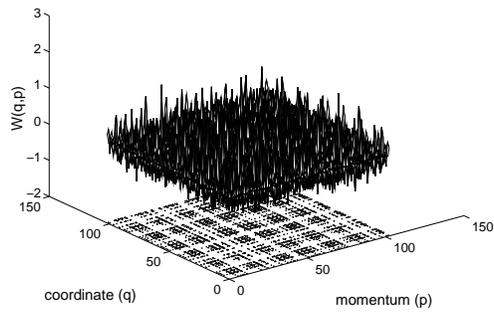}
\caption{\bf Entangled-like Wigner function.}
\end{figure}

\begin{figure}[ht]
\centering
\includegraphics*[width=65mm]{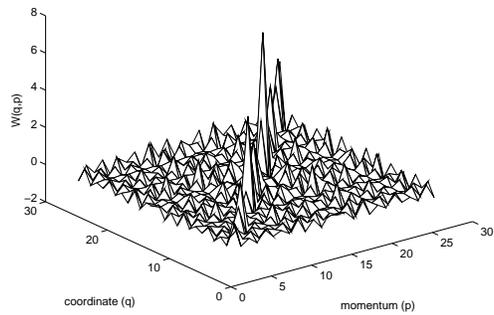}
\caption{\bf Localized pattern: waveleton Wigner function.}
\end{figure}

\end{document}